\documentstyle[sprocl]{article}

\def\be{\begin{equation}}
\def\ee{\end{equation}}
\def\bea{\begin{eqnarray}}
\def\eea{\end{eqnarray}}

\newcommand{\qrd}[2]{{\em Phys. Rev.}               {\bf D#1}, #2 }
\newcommand{\qrl}[2]{{\em Phys. Rev. Lett.}         {\bf  #1}, #2 }
\newcommand{\zpc}[2]{{\em Z. Phys.}                 {\bf C#1}, #2 }

\newcommand{\etal}{{\em et al.}}

\newcommand{\col}{Collaboration}
\newcommand{\ba}{\begin{array}}
\newcommand{\ea}{\end{array}}
\newcommand{\ms}{{\overline{\rm MS}}}

\begin{document}

\title{Implications of Precision Electroweak Measurements
for the Standard Model Higgs Boson\footnote{Talk presented at the
17th International Workshop on Weak Interactions and
Neutrinos (WIN99), Cape Town, South Africa, January 24--30, 1999.}}

\author{J. ERLER}

\address{Department of Physics and Astronomy, University of Pennsylvania,
\\ Philadelphia, PA 19104-6396, USA \\ E-mail: erler@langacker.hep.upenn.edu} 

\maketitle
\abstracts{We summarize the status of the Standard Model with special 
emphasis on the extraction of the Higgs boson mass using Bayesian inference.}

\section{Introduction}
Besides the recent high precision measurements of the $W$ 
mass~\cite{Karlen98,Dorigo98}, $M_W$, the most important input into precision 
tests of electroweak theory continues to come from the $Z$ factories 
LEP~1~\cite{Karlen98} and SLC~\cite{Baird98}. The vanguard of the physics 
program at LEP~1 is the analysis of the $Z$ lineshape. Its parameters are the 
$Z$ mass, $M_Z$, the total $Z$ width, $\Gamma_Z$, the hadronic peak cross 
section, $\sigma_{\rm had}$, and the ratios of hadronic to leptonic decay 
widths, $R_\ell = {\Gamma({\rm had})\over \Gamma(\ell^+\ell^-)}$, where 
$\ell = e$, $\mu$, or $\tau$. They are determined in a common fit with the 
leptonic forward-backward (FB) asymmetries, 
$A_{FB} (\ell) = {3\over 4} A_e A_\ell$. With $f$ denoting the fermion index, 
\be
  A_f = {2 v_f a_f\over v_f^2 + a_f^2}
\ee
is defined in terms of the vector 
($v_f = I_{3,f} - 2 Q_f \sin^2 \theta_f^{\rm eff}$) and axial-vector 
($a_f = I_{3,f}$) $Zf\bar{f}$ coupling; $Q_f$ and $I_{3,f}$ are the electric 
charge and third component of isospin, respectively, and 
$\sin^2 \theta_f^{\rm eff} \equiv \bar{s}^2_f$ is an effective mixing angle.

The polarization of the electron beam at the SLC allows for
competitive and complementary measurements with a much smaller 
number of $Z$'s than at LEP. In particular, the left-right (LR) cross 
section asymmetry, $A_{LR} = A_e$, represents the most precise determination 
of the weak mixing angle by a single experiment (SLD).~\cite{Baird98} 
Mixed FB-LR asymmetries, $A^{FB}_{LR} (f) = {3\over 4} A_f$, single out the 
final state coupling of the $Z$ boson. 

For several years there has been an experimental discrepancy at the $2 \sigma$ 
level between $A_\ell$ from LEP and the SLC. With the 1997/98 high statistics 
run at the SLC, and a revised value for the FB asymmetry of the $\tau$ 
polarization, ${\cal P}^{FB}_\tau$, the two determinations are now consistent 
with each other,
\be \ba{l}
\label{aell}
  A_\ell ({\rm LEP}) = 0.1470 \pm 0.0027, \\
  A_\ell ({\rm SLD}) = 0.1503 \pm 0.0023.
\ea \ee

\begin{table}[p]
\caption{Principal precision observables from CERN, FNAL, SLAC, and elsewhere.
Shown are the experimental results, the SM predictions, and the pulls. 
The SM errors are from the 
uncertainties in $M_Z$, $\ln M_H$, $m_t$, $\alpha (M_Z)$, and $\alpha_s$. 
They have been treated as Gaussian and their correlations have been taken into 
account. $\bar{s}_\ell^2 (Q_{FB} (q))$ is the weak mixing angle from the 
hadronic charge asymmetry; $R^-$ and $R^\nu$ are cross section ratios from 
deep inelastic $\nu$-hadron scattering; $g_{V,A}^{\nu e}$ are effective 
four-Fermi coefficients in $\nu$-e scattering; and the $Q_W$ are the weak 
charges from parity violation measurements in atoms. The uncertainty in the 
$b\rightarrow s\gamma$ observable includes theoretical errors from the physics
model, the finite photon energy cut-off, and from uncalculated higher order 
effects. There are other precision observables which are not shown but 
included in the fits. Very good agreement with the SM is observed. 
Only $A_{LR}$ and the two measurements sensitive to $A_b$ discussed in the
text, show some deviation, but even those are below $2\sigma$.
\label{zpole}}
\vspace{0.2cm}
\begin{center}
\footnotesize
\begin{tabular}{|lcccr|}
\hline
Quantity & Group(s) & Value & Standard Model & pull \\ 
\hline
$M_Z$       \hspace{0pt} [GeV]&     LEP     &$ 91.1867 \pm 0.0021 $&$ 91.1865 \pm 0.0021 $&$ 0.1$ \\
$\Gamma_Z$  \hspace{3pt} [GeV]&     LEP     &$  2.4939 \pm 0.0024 $&$  2.4957 \pm 0.0017 $&$-0.8$ \\
$\sigma_{\rm had}$       [nb] &     LEP     &$ 41.491  \pm 0.058  $&$ 41.473  \pm 0.015  $&$ 0.3$ \\
$R_e$                         &     LEP     &$ 20.783  \pm 0.052  $&$ 20.748  \pm 0.019  $&$ 0.7$ \\
$R_\mu$                       &     LEP     &$ 20.789  \pm 0.034  $&$ 20.749  \pm 0.019  $&$ 1.2$ \\
$R_\tau$                      &     LEP     &$ 20.764  \pm 0.045  $&$ 20.794  \pm 0.019  $&$-0.7$ \\
$A_{FB} (e)$                  &     LEP     &$  0.0153 \pm 0.0025 $&$  0.0161 \pm 0.0003 $&$-0.3$ \\
$A_{FB} (\mu)$                &     LEP     &$  0.0164 \pm 0.0013 $&$                    $&$ 0.2$ \\
$A_{FB} (\tau)$               &     LEP     &$  0.0183 \pm 0.0017 $&$                    $&$ 1.3$ \\
\hline
$R_b$                         &  LEP + SLD  &$  0.21656\pm 0.00074$&$  0.2158 \pm 0.0002 $&$ 1.0$ \\
$R_c$                         &  LEP + SLD  &$  0.1735 \pm 0.0044 $&$  0.1723 \pm 0.0001 $&$ 0.3$ \\
$A_{FB} (b)$                  &     LEP     &$  0.0990 \pm 0.0021 $&$  0.1028 \pm 0.0010 $&$-1.8$ \\
$A_{FB} (c)$                  &     LEP     &$  0.0709 \pm 0.0044 $&$  0.0734 \pm 0.0008 $&$-0.6$ \\
$A_b$                         &     SLD     &$  0.867  \pm 0.035  $&$  0.9347 \pm 0.0001 $&$-1.9$ \\
$A_c$                         &     SLD     &$  0.647  \pm 0.040  $&$  0.6676 \pm 0.0006 $&$-0.5$ \\
\hline
$A_{LR} + A_\ell$             &     SLD     &$  0.1503 \pm 0.0023 $&$  0.1466 \pm 0.0015 $&$ 1.6$ \\
${\cal P}_\tau: A_e+A_\tau$   &     LEP     &$  0.1452 \pm 0.0034 $&$                    $&$-0.4$ \\
$\bar{s}_\ell^2 (Q_{FB})$     &     LEP     &$  0.2321 \pm 0.0010 $&$  0.2316 \pm 0.0002 $&$ 0.5$ \\
\hline
$m_t$      \hspace{6pt} [GeV]&  Tevatron   &$173.8    \pm 5.0    $&$171.4    \pm 4.8    $&$ 0.5$ \\
$M_W$      \hspace{0pt}  [GeV]&     all     &$ 80.388  \pm 0.063  $&$ 80.362  \pm 0.023  $&$ 0.4$ \\
\hline
$R^-$          &     NuTeV      &$   0.2277 \pm 0.0021 \pm 0.0007 $&$   0.2297 \pm 0.0003 $&$-0.9$\\
$R^\nu$        &     CCFR       &$   0.5820 \pm 0.0027 \pm 0.0031 $&$   0.5827 \pm 0.0005 $&$-0.2$\\
$R^\nu$        &     CDHS       &$   0.3096 \pm 0.0033 \pm 0.0028 $&$   0.3089 \pm 0.0003 $&$ 0.2$\\
$R^\nu$        &     CHARM      &$   0.3021 \pm 0.0031 \pm 0.0026 $&$                     $&$-1.7$\\
\hline
$g_V^{\nu e}$  &      all       &$  -0.041  \pm 0.015             $&$  -0.0395 \pm 0.0004 $&$-0.1$\\
$g_A^{\nu e}$  &      all       &$  -0.507  \pm 0.014             $&$  -0.5063 \pm 0.0002 $&$-0.1$\\
\hline
$Q_W({\rm Cs})$&     Boulder    &$ -72.41   \pm 0.25\pm 0.80      $&$ -73.10   \pm 0.04   $&$ 0.8$\\
$Q_W({\rm Tl})$&      all       &$-114.8    \pm 1.2 \pm 3.4       $&$-116.7    \pm 0.1    $&$ 0.5$\\
\hline
${\Gamma (b\rightarrow s\gamma)\over \Gamma (b\rightarrow c e\nu)}$& CLEO 
           &$ 3.26^{+0.75}_{-0.68} \times 10^{-3} $&$ 3.14^{+0.19}_{-0.18} 
           \times 10^{-3} $&$ 0.1$\\
\hline
\end{tabular}
\end{center}
\end{table}

\noindent
The LEP value is from $A_{FB}(\ell)$, ${\cal P}_\tau$, and 
${\cal P}^{FB}_\tau$, while the SLD value is from $A_{LR}$ and 
$A^{FB}_{LR} (\ell)$. The data is consistent with lepton universality, 
which is assumed here. There remains a $2.5 \sigma$ discrepancy between 
the two most precise determinations of $\bar{s}^2_\ell$, i.e.\ $A_{LR}$ 
and $A_{FB} (b)$ (assuming no new physics in $A_b$).

Of particular interest are the results on the heavy flavor 
sector~\cite{Karlen98} including 
$R_q = {\Gamma (q\bar{q}) \over \Gamma ({\rm had})}$, $A_{FB} (q)$, and 
$A^{FB}_{LR} (q)$, with $q = b$ or $c$. At present, there is some
discrepancy in $A^{FB}_{LR} (b) = {3\over 4} A_b$ and 
$A_{FB} (b) = {3\over 4} A_e A_b$, both at the $2 \sigma$ level. Using 
the average of Eqs.~(\ref{aell}), $A_\ell = 0.1489 \pm 0.0018$, both can be 
interpreted as measurements of $A_b$. From $A_{FB} (b)$ one would obtain 
$A_b = 0.887 \pm 0.022$, and the combination with 
$A^{FB}_{LR} (b) = {3\over 4} (0.867 \pm 0.035)$ would yield 
$A_b = 0.881 \pm 0.019$, which is almost $3 \sigma$ below the SM prediction. 
Alternatively, one could use $A_\ell ({\rm LEP})$ above (which is closer to the
SM prediction) to determine $A_b ({\rm LEP}) = 0.898 \pm 0.025$, and 
$A_b = 0.888 \pm 0.020$ after combination with $A^{FB}_{LR} (b)$, i.e., still 
a $2.3 \sigma$ discrepancy. An explanation of the 5--6\% deviation in $A_b$
in terms of new physics in loops, would need a 25--30\% radiative correction 
to $\hat\kappa_b$, defined by 
$\bar{s}^2_b \equiv \hat\kappa_b\sin^2\hat\theta_\ms (M_Z)$.
Only a new type of physics which couples at the tree level 
preferentially to the third generation~\cite{Erler95}, and which does not 
contradict $R_b$ (including the off-peak measurements by 
DELPHI~\cite{Abreu96}), can conceivably account for a low $A_b$. 
Given this and that none of the observables deviates by $2 \sigma$ or more, 
we can presently conclude that there is no compelling evidence for new physics 
in the precision observables, some of which are listed in Table~\ref{zpole}. 

\section{Bayesian Higgs mass inference}

The data show a strong preference for a low $M_H \sim {\cal O} (M_Z)$,
\be
\label{mh_fit}
   M_H = 107^{+67}_{-45} \mbox{ GeV},
\ee
where the central value (of the global fit to all precision data, including 
$m_t$) maximizes the likelihood, $N e^{-\chi^2 (M_H)/2}$. Correlations with 
other parameters, $\xi^i$, are accounted for, since minimization w.r.t. these 
is understood, $\chi^2 \equiv \chi^2_{\rm min}$. 

Bayesian methods, on the other hand, are based on Bayes theorem~\cite{Bayes63},
\be
\label{Bayes}
  p(M_H | {\rm data}) = \frac{p({\rm data}| M_H) p(M_H)}{p({\rm data})},
\ee
which must be satisfied once the {\em likelihood\/}, $p({\rm data}| M_H)$, and 
{\em prior\/} distribution, $p(M_H)$, are specified.
$p(data) \equiv \int p({\rm data}| M_H) p(M_H) d M_H$ in the denominator
provides for the proper normalization of the {\em posterior\/} distribution on
the l.h.s. The prior can contain additional information not included in the 
likelihood model, or chosen to be {\em non-informative}. 

Occasionally, the Bayesian method is criticized for the need of  
a prior, which would introduce unnecessary subjectivity into the 
analysis. Indeed, care and good judgement is needed, but the same is true 
for the likelihood model, which has to be specified in both approaches.
Moreover, it is appreciated among Bayesian practitioners, that the explicit 
presence of the prior can be advantageous: it manifests model assumptions 
and allows for sensitivity checks. From the theorem~(\ref{Bayes}) it is 
also clear that the maximum likelihood method corresponds, mathematically, 
to a particular choice of prior. Thus Bayesian methods differ rather 
in attitude: by their strong emphasis on the entire posterior distribution 
and by their first principles setup. 

Given extra parameters, $\xi^i$, the distribution function of $M_H$ is defined 
as the marginal distribution, 
$p(M_H|{\rm data}) = \int p(M_H, \xi^i | {\rm data}) \prod_i p(\xi^i) d \xi^i$.
If the posterior factorizes, $p(M_H, \xi^i) = p(M_H) p(\xi^i)$, the $\xi^i$ 
dependence can be ignored. If not, but $p(\xi^i | M_H)$ is 
(approximately) multivariate normal, then
\be
  \chi^2 (M_H,\xi^i) = \chi^2_{\rm min} (M_H) +
  {1\over 2} \frac{\partial^2 \chi^2 (M_H)} {\partial \xi_i \partial \xi_j} 
  (\xi^i - \xi^i_{\rm min} (M_H)) (\xi^j - \xi^j_{\rm min} (M_H)).
\ee
The latter applies to our case, where $\xi^i = (m_t,\alpha_s,\alpha(M_Z))$. 
Integration yields, 
\be
  p(M_H | {\rm data}) \sim \sqrt{\det E}\, e^{- \chi^2_{\rm min} (M_H)/2},
\ee
where the $\xi^i$ error matrix, $E = (\frac{\partial^2 \chi^2 (M_H)}
  {\partial \xi_i \partial \xi_j})^{-1}$, introduces a correction factor
with a mild $M_H$ dependence. It corresponds to a shift relative to the 
standard likelihood model, 
$\chi^2 (M_H) = \chi^2_{\rm min}(M_H) + \Delta \chi^2 (M_H)$, where
\be
  \Delta \chi^2 (M_H) \equiv \ln \frac{\det E (M_H)}{\det E (M_Z)}.
\ee
For example, $\Delta \chi^2 (300 \mbox{ GeV}) \sim 0.1$, which would 
{\em tighten} the $M_H$ upper limit by at most a few GeV. At present, 
we neglect this effect.
 
We choose $p(M_H)$ as the product of $M_H^{-1}$, corresponding to a uniform
(non-informative) distribution in $\log M_H$, times the exclusion curve from 
LEP~2.~\cite{McNamara98} This curve is from Higgs searches at center of mass 
energies up to 183 GeV. We find the 90 (95, 99)\% confidence upper limits,
\be
\label{mh_limits}
  M_H < 220 \mbox{ (255, 335) GeV}.
\ee
Theory uncertainties from uncalculated higher orders increase the 95\% CL 
by about 5~GeV. These limits are robust within the SM, but we caution that 
the results on $M_H$ are strongly correlated with certain new physics 
parameters~\cite{Erler99}.

The one-sided confidence interval~(\ref{mh_limits}) is not an exclusion limit.
For example, the 95\% upper limit of the standard uniform distribution, 
$x \in [0,1]$, is at $x = 0.95$, but all values of $x$ are equally likely, and 
$x > 0.95$ cannot be excluded. If there is a discrete set of competing
hypotheses, $H_i$, one can use Bayes factors, 
$p({\rm data} | H_i)/p({\rm data} | H_j)$, for comparison. For example, LEP~2 
rejects a standard Higgs boson with $M_H < 90$~GeV at the 95\% CL, because
\be 
   \frac{p({\rm data} | M_H   =  M_0)}{p({\rm data} | M_H \neq M_0)} 
   < 0.05 \hspace{20pt} \forall\; M_0 < 90 \hbox{ GeV}.
\ee
On the other hand, 
the probability for $M_H < 90$~GeV is only $5\times 10^{-4}$. 

One could similarly note, that
$p(M_H = M_0) < 0.05\, p(M_H = 107 \hbox{ GeV})$ for $M_0 > 334$ GeV; 
but the (arbitrary) choice of the best fit $M_H$ value as reference 
hypothesis is hardly justifiable. This affirms that variables
continuously connecting a set of hypotheses should be treated in a fully 
Bayesian analysis.

\section*{Acknowledgement}
I would like to thank the organizers of WIN 99 for a very pleasant and 
memorable meeting and Paul Langacker for collaboration.


\section*{References}
\begin{thebibliography}{99}
\bibitem{Karlen98}
  D. Karlen, {\em Experimental Status of the Standard Model},
  Talk presented at the XXIXth International Conference on High Energy Physics
  (ICHEP~98), Vancouver, Canada, July 1998; \\
  The LEP Collaborations ALEPH, DELPHI, L3, OPAL, the LEP Electroweak Working 
  Group, and the SLD Heavy Flavour and Electroweak Groups: 
  D. Abbaneo \etal, Internal Note CERN--EP/99--15.
\bibitem{Dorigo98}
  T. Dorigo for the CDF \col: {\em Electroweak Results from the Tevatron 
  Collider}, Talk presented at PASCOS 98; \\
  D\O\ \col : B. Abbott \etal, \qrl{80}{3008}(1998). 
\bibitem{Baird98} 
  K. Baird for the SLD \col: 
  {\em Measurements of $A_{LR}$ and $A_{\rm lepton}$ from SLD},
  Talk presented at ICHEP~98.
\bibitem{Erler95}
  J. Erler, \qrd{52}{28}(1995); \\
  J. Erler, J.L. Feng, and N. Polonsky, \qrl{78}{3063}(1997).
\bibitem{Abreu96}
  DELPHI \col: P. Abreu \etal, \zpc{70}{531}(1996).
\bibitem{Bayes63}
   T. Bayes, {\em Phil. Trans.} {\bf 53}, 370 (1763) and 
   {\em Biometrica} {\bf 45}, 296 (1958).
\bibitem{McNamara98}
  P. McNamara, {\em Standard Model Higgs at LEP},
  talk presented at ICHEP~98.
\bibitem{Erler99}
  J. Erler, {\em Implications of Precision Electroweak Measurements
  for Physics Beyond the Standard Model}, these proceedings.
\end{thebibliography}
\end{document}